\documentclass[letter]{aa}

\usepackage{graphicx}
\usepackage{natbib}
\usepackage{mathrsfs}

\newcommand{\kms}{km~s$^{-1}$}

\newcommand{\dg}{$^{\circ}$}
\newcommand{\rsun}{$R_{\odot}$}
\newcommand{\xiv}{~{\sc{xiv}}}

\newcommand{\tii}{type~{\sc{ii}}}
\newcommand{\alfven}{Alfv\'en}
\newcommand{\alfvenic}{Alfv\'enic}

\begin{document}

\title{Three-dimensional reconstruction of CME-driven shock--streamer interaction from radio and EUV observations: \\ a different take on the diagnostics of coronal magnetic fields}

\author{S. Mancuso\inst{1}, F. Frassati\inst{1,2}, A. Bemporad\inst{1}, \and D. Barghini\inst{1,2}}

\institute{
${^1}$ Istituto Nazionale di Astrofisica, Osservatorio Astrofisico di Torino, via Osservatorio 20, Pino T.se 10025, Italy \\ 
${^2}$ Universit\`a degli Studi di Torino, Dipartimento di Fisica, via Pietro Giuria 1, Torino (TO), Italy \\ \email{salvatore.mancuso@inaf.it}}
\date{Received / Accepted}

\abstract{On 2014 October 30, a band-splitted \tii\ radio burst associated with a coronal mass ejection (CME) observed by the Atmospheric Imaging Assembly (AIA) on board the Solar Dynamic Observatory (SDO) occurred over the southeast limb of the Sun.
The fast expansion in all directions of the plasma front acted as a piston and drove a spherical fast shock ahead of it, whose outward progression was traced by simultaneous images obtained with the Nan\c{c}ay Radioheliograph (NRH).
The geometry of the CME/shock event was recovered through 3D modeling, given the absence of concomitant stereoscopic observations, and assuming that the band-splitted \tii\ burst was emitted at the intersection of the shock surface with two adjacent low-\alfven\ speed coronal streamers.
From the derived spatiotemporal evolution of the standoff distance between shock and CME leading edge, we were finally able to infer the magnetic field strength $B$ in the inner corona. 
A simple radial profile of the form $B(r) = (12.6 \pm 2.5) r^{-4}$ nicely fits our results, together with previous estimates, in the range $r = 1.1-2.0$ \rsun.
}
\titlerunning{3D reconstruction of CME-driven shock--streamer interaction from radio and EUV observations}
\authorrunning{Mancuso et al.}
\maketitle

\section{Introduction}

Determining the strength of the coronal magnetic field remains a central problem in solar physics.
In situ measurements in the higher corona will be available with the Solar Probe Plus mission, launched in August 2018, but direct information on the magnetic field below about 8 solar radii (\rsun) will probably remain out of reach for decades.
Direct measurements of magnetic field strengths $B$ are extremely difficult to make and are restricted to the bottom of the coron
a (e.g., \citealt{Aurass1987,Lin2000,Nakariakov2001,Ofman2008,Verwichte2009}).
Indirect techniques, such as Faraday rotation measurements of occulted extragalactic radio sources, represent an alternative approach in estimating $B$ but only above about 5 \rsun\ (e.g., \citealt{Mancuso2000,Spangler2005,Mancuso2013a,Mancuso2013b}) due to instrumental limitations.
However, estimates at just a few \rsun\ are of pivotal importance since the strength and structure of the magnetic field in the inner corona crucially affects the acceleration of the solar wind, which in turn has strong influence on space weather.
Among the techniques for measuring coronal magnetic fields, suitable diagnostics are provided by the analysis of remote sensing coronagraphic observations acquired during the passage of coronal shock waves associated to coronal mass ejections (CMEs; see review by \citealt{Bemporad2016}).

In this Letter, we use a technique developed by \cite{Gopalswamy2011}, applied to a CME-driven shock observed on 2014 October 30, which involves measuring the shock standoff distance and the radius of curvature of the driving CME as deduced from both near-Sun extreme ultraviolet (EUV) and radioheliograph images to derive the radial profile $B(r)$ of the upstream magnetic field strength. 
This result will be obtained, in an unprecedented way, through 3D modeling without the use of stereoscopic observations but only by exploiting the peculiar band-splitting of the associated \tii\ radio burst whose sources are found to be excited at the intersections of the expanding shock surface with two adjacent coronal streamers.

\section{Observations and data analysis}

\begin{figure}[t]
\centering
\hspace{-0.1cm}
\vspace{-0.1cm}
\includegraphics[width=3.7cm]{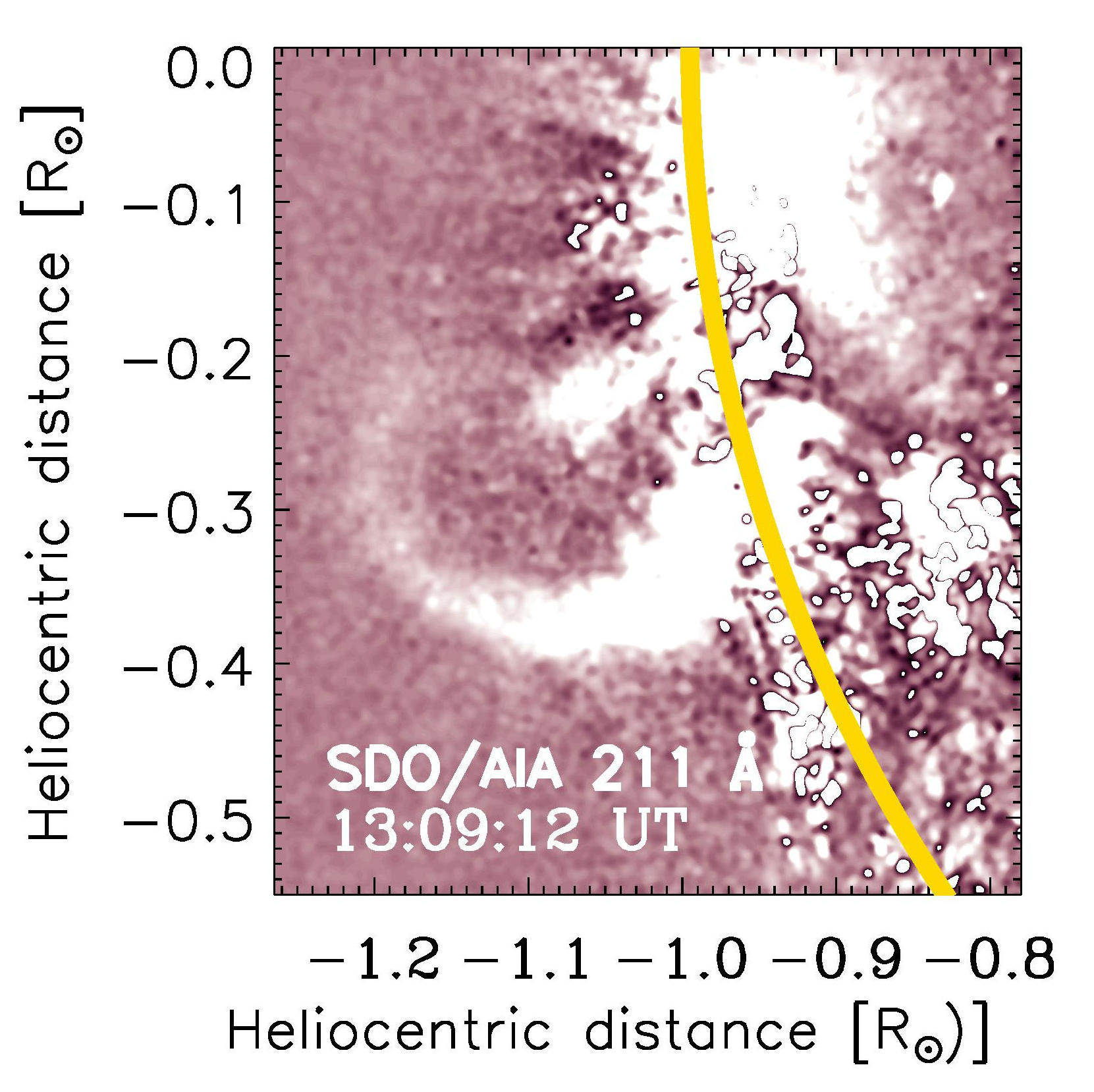}
\hspace{+0.5cm}
\vspace{+0.2cm}
\includegraphics[width=3.6cm]{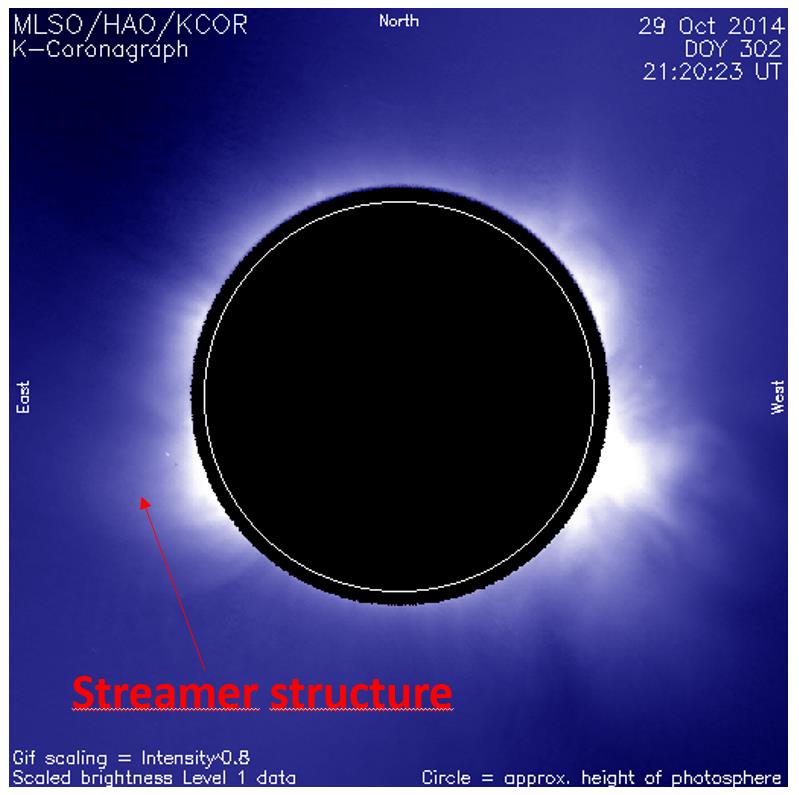}
\caption{
{\it Left panel}: AIA/SDO base difference EUV images of the eruption at 211~\AA. 
A bubble-like, hemispherical expanding EUV front is clearly visible from the images taken at 13:09:12 UT.
The yellow curve identifies the surface of the Sun.
{\it Right panel}: MLSO K-Cor white-light image taken on the day before the event.}
\label{Fig1}
\end{figure}

\begin{figure}[t]
\centering
\includegraphics[width=8.2cm]{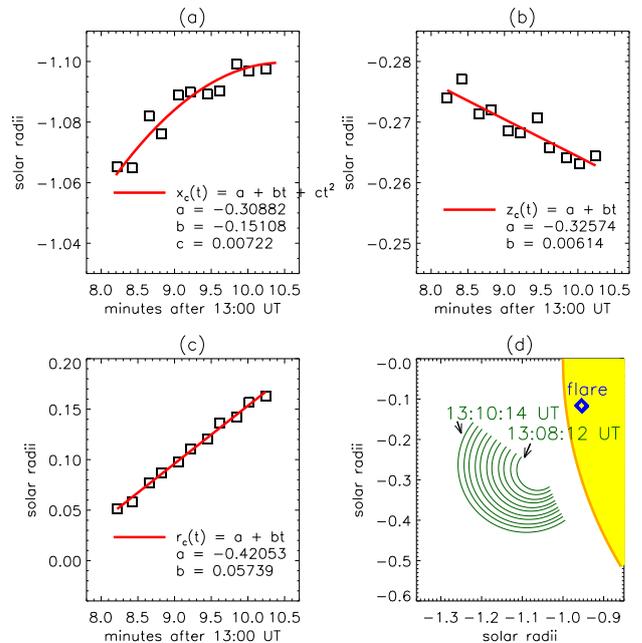}
\caption{
Quadratic fits to the temporal evolution of the three parameters $x_{\rm c}$ ({\it panel a}), $z_{\rm c}$ ({\it panel b}), and $r_{\rm cme}$ ({\it panel c}) representing, respectively, the $x$ and $z$ coordinates in the plane of the sky $x-z$ of the center and the radius of the circle ({\it panel d}) used to fit the bright expanding bubble observed by SDO/AIA between 13:08:12 UT and 13:10:14 UT.}
\label{Fig2}
\end{figure}

On 2014 October 30, a solar eruption occurred near the southeast limb above active region NOAA 12201 (S04E70) in association with a C6.9 flare.
The early evolution of the CME was observed by the Atmospheric Imaging Assembly (AIA;~\citealt{Lemen2012}) on board the Solar Dynamic Observatory (SDO) in the EUV bandpass centered on Fe\xiv\ (211 \AA) corresponding to coronal plasma at temperatures of $\sim1.6$ MK.
AIA difference images (Fig.~\ref{Fig1}, left panel) reveal a sharp expanding loop-shaped coronal EUV front structure which eventually evolved into a CME observed above 2 \rsun\ with white-light coronagraphs.
The EUV front seen by AIA is interpreted here as the outermost layer of compressed and piled-up plasma moving through the corona, acting as the contact surface that creates the shock wave ahead of it.
The observed bright EUV rim could be produced in principle by an expanding face-on quasi-circular loop. 
Alternatively, it could represent the projection of a bubble-like structure on the plane of the sky.
Unfortunately, no stereoscopic observations are available for this event to validate one or the other hypothesis.
Using stereoscopic data, \cite{Patsourakos2010} presented a detailed analysis of the formation and early evolution phase of an impulsively accelerated CME. 
They showed that the CME formed a rapidly expanding bubble appearing as a nearly circular bright rim in EUV images but that was actually the projection of a 3D structure that could be described reasonably well by a sphere in its early development stages. 
From the analysis of the high-cadence AIA images of this event, given the nearly circular shape of the expanding EUV front, we choose to adopt this latter interpretation and thus model the evolution of the EUV front as an expanding spherical bubble with its center translating in space with time.
To validate the above assumption, we show that the chosen 3D geometry nicely explains the evolution of the observed radio sources linked to the shock front development in the inner corona.
The dynamics of the expanding surface was obtained by fitting to the data the temporal evolution of the coordinates in the plane of the sky $x-z$ of the center of the circle [$x_{\rm c}(t),z_{\rm c}(t)$] and of its radius $r_{\rm cme}(t)$ with low-order polynomials (Fig.~\ref{Fig2}).

\begin{figure}[t]
\centering
\includegraphics[width=8.2cm]{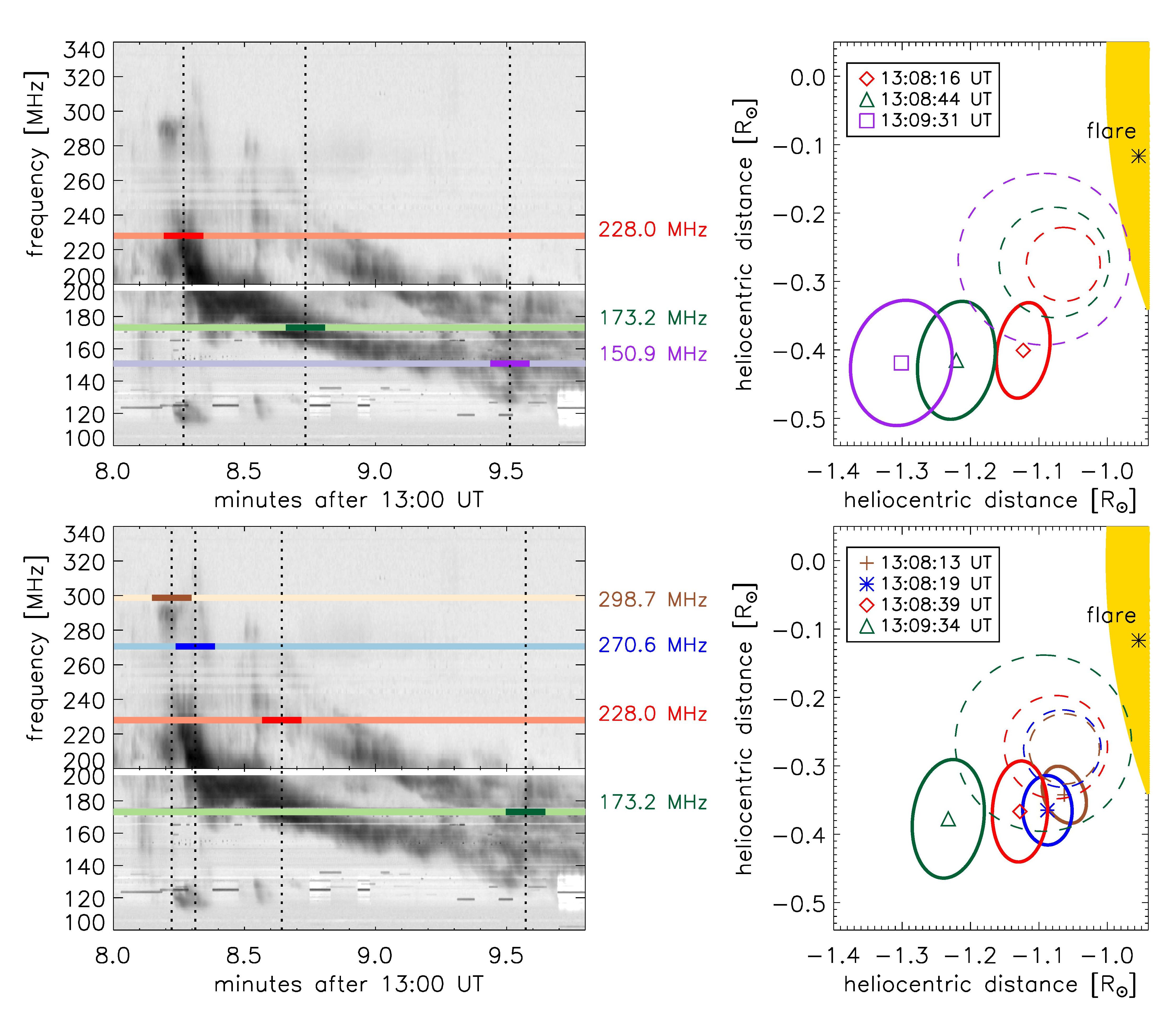}
\caption{
{\it Left panels}: Combined {\it mid} ($100-196$ MHz) and {\it high} ($200-340$ MHz) dynamic spectra obtained by BIR CALLISTO receivers covering the high-frequency range of the \tii\ event detected on 2014 October 30. 
Darker-colored lines show the time intervals over which the NRH data have been averaged at a given frequency.
{\it Right panels}: Locations of the centroids of 2D elliptical Gaussian fits (shown with 95\% contour levels) to the \tii\ sources observed by NRH (plotted with colors corresponding to the frequencies in the left panels).
Dashed circles correspond to the corresponding EUV front locations.}
\label{Fig3}
\end{figure}

A \tii\ radio burst, indicating the presence of a shock wave, was also detected by ground-based radio spectrographs.
Figure~\ref{Fig3} displays the radio dynamic spectrum obtained during the event by the Callisto radio spectrometer (\citealt{Benz2005}) network station at the Birr castle in Ireland. 
After 13:08 UT, the \tii\ radio emission lanes stand out clearly, showing a weak fundamental emitted at the electron plasma frequency $f_\mathrm{pe} \approx 9 \sqrt{n_\mathrm{e}[\rm cm^{-3}]}$ kHz at about 120 MHz, and a much stronger, splitted harmonic band at higher frequencies.
The harmonic component is split into two sub-bands, a lower (L) and an upper (U) frequency component.
The \tii\ radio sources were also imaged by the Nan\c{c}ay Radioheliograph (NRH;~\citealt{Kerdraon1997}) at a time that was particularly favorable for making precise observations because of the small zenith distance of the Sun.
The approximate locations of the centroids of the NRH sources associated with the \tii\ burst were estimated by fitting 2D elliptical Gaussian functions to each of the NRH images.
We were thus able to localize the L and U frequency sources at five NRH frequencies (298.7, 270.6, 228.0, 173.2, and 150.9 MHz; see Fig.~\ref{Fig3}) and thus retrieve the electron density as a function of space and time, information that is important to infer the magnetic field strength at those heights.
Since the NRH frequencies are much higher than the ionospheric cut-off frequency (a few tens of MHz), refraction effects on determining the location of the radio source are not important here (\citealt{Riddle1974}).

The origin of band-splitting in \tii\ radio bursts is still controversial.
One possible explanation first proposed by \citet{Smerd1975} is that the two lanes originate from simultaneous radio emission occurring in the upstream (ahead) and downstream (behind) regions of a shock.
In this scenario, the radio sources emitting the two splitted frequencies near the Sun should be virtually cospatial as the shock thickness is negligible.
Recently, using observations from the LOw-Frequency ARray (LOFAR; \citealt{vanHarlem2013}), \cite{Chrysaphi2018} and \cite{Zucca2018} found that the sources of the two split bands in their events were cospatial, in agreement with the interpretation that split bands are due to simultaneous emission from plasma upstream and downstream of the shock front.
Since NRH data sample only a few frequencies and the observations outlined in Fig.~\ref{Fig3} are not strictly simultaneous for the L and U branches of the splitted \tii\ emissions, it is not straightforward to discount simultaneous cospatial L and U emissions even for this event.
However, the centroids of maximum intensity in each band for the almost simultaneous observations at 13:09:31 UT (at 150.9 MHz) taken in the L band and at 13:09:34 UT (at 173.2 MHz) taken in the U band are clearly not cospatial, as also the centroids of the almost simultaneous observations taken in the U band at 13:08:13 UT (at 298.7 MHz) and 13:08:19 UT (at 270.6 MHz) and in the L band at 13:08:16 UT (at 228.0 MHz).
Notwithstanding the above, if we consider that the minor (major) axes of the ellipses that define the half-power beamwidth at 13:08 UT range 
from 0.067 (0.129) \rsun\ at 298.7 MHz to 0.133 (0.256) \rsun\ at 150.9 MHz, cospatiality of band-split emissions, although unlikely, cannot be completely ruled out. For comparison, the difference between the centroid locations in the east-west direction of the radio sources seen almost simultaneously in the U band at 173.2 MHz and in the L band at 150.9 MHz is about 0.07 \rsun. 
However, as stated above, also considering that refraction effects are not important here, an alternative explanation at least for this event seems more plausible.

\begin{figure}[t]
\centering
\includegraphics[width=7.8cm]{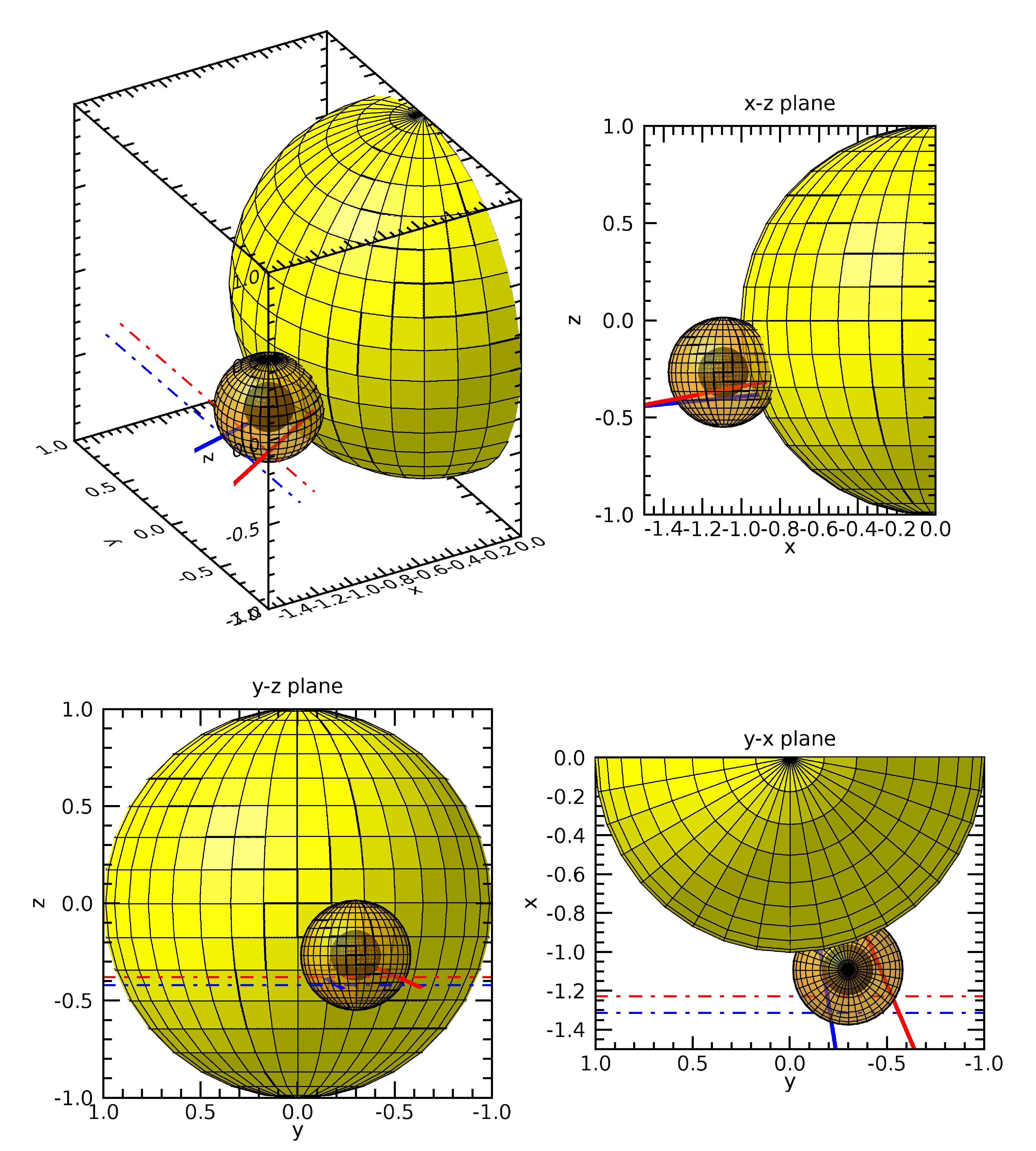}
\caption{
Three-dimensional model of the CME/shock surface (here shown at 13:09:30 UT). 
The $x-z$ plane identifies the plane of the sky, where $z$ is the north direction. 
The \tii\ emission is supposed to be excited at the intersection of the spherical shock surface (identified by the light-brown sphere surrounding the dark-brown CME bubble) and the streamer axes. 
Blue and red straight lines identify the axes of the streamers and correspond, respectively, to \tii\ emissions in the lower (L) and upper (U) branches of the splitted \tii. 
Two lines of sight are depicted as red and blue lines with dots. 
The model explains the features observed by NRH at the various frequencies in Fig.~\ref{Fig3}.}
\label{Fig4}
\end{figure}

\begin{figure*}
\centering
\includegraphics[width=16cm]{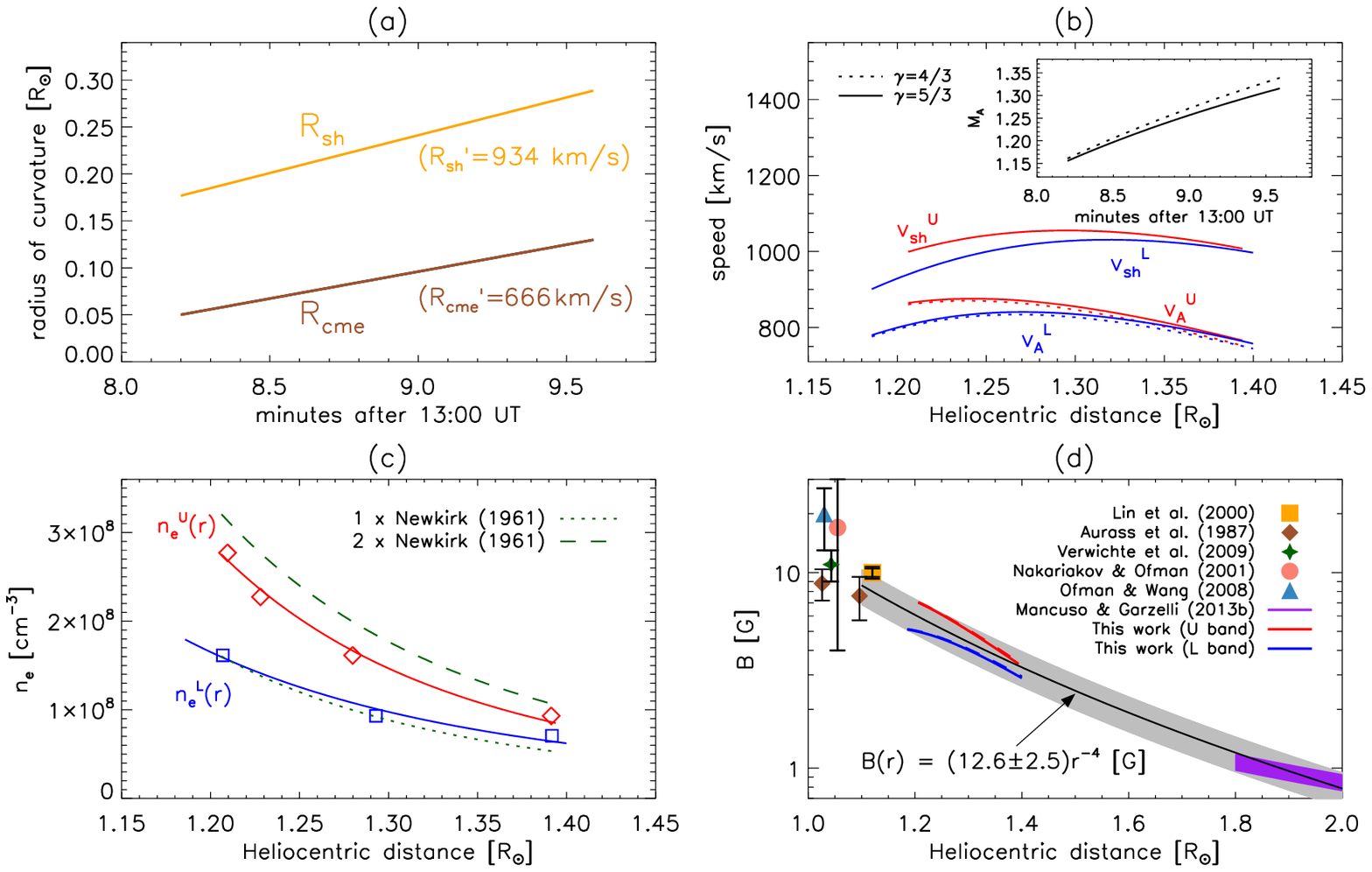}
\caption{
{\it Panel a}: Radial profiles of the radii of curvature of the shock and the CME leading edge.
{\it Panel b}: Radial profiles of $v_{\rm sh}$ and $v_{\rm A}$ along the two streamers depicted in Fig.~\ref{Fig4} and time profile of $M_{\rm A}$ (see the inset). Red and blue colors correspond to the colors of the streamer axes in that figure.
The U (L) suffix highlights the streamer intersected by the shock surface emitting at the upper (lower) frequency band of the \tii\ shown in Fig.~\ref{Fig3}.
{\it Panel c}: Radial profiles of $n_{\rm e}$ along the two streamers depicted in Fig.~\ref{Fig4} compared with one- and two-fold \citet{Newkirk1961} radial profiles.
{\it Panel d}: Magnetic-field radial profile as derived in this work (red and blue curves), together with estimates of $B$ based on different methods.
A simple radial profile (black line within gray boundaries) nicely fits our results and the previous estimates in the range $1.1-2.0$ \rsun.}
\label{Fig5}
\end{figure*}

Band-splitting can also be generated more simply by emission from two different parts of the same shock front expanding through plasma structures with different electron density and magnetic field distributions.
In fact, the observed radio sources evolve nonradially, with the \tii\ emissions confined in a small portion of the corona and aligned in a well-defined direction.
This suggests that the \tii\ sources likely originate at the intersection of the shock surface with enhanced density (and thus low-\alfven\ speed) regions of the corona, more probably in coronal streamers.
In fact, the strength of fast-mode shocks in the corona can vary because of the uneven distribution of $v_\mathrm{A}$ but is very much enhanced on those parts of the wave front that encounter low-$v_\mathrm{A}$ structures such as streamers, as suggested by \cite{Mancuso2004}. 
The NRH observations shown in Fig.~\ref{Fig3} would therefore represent \tii\ emission at the intersection of the expanding shock surface with the axes of the streamers. 
As a matter of fact, inspection of Mauna Loa Solar Observatory (MLSO) white light K-coronagraph (K-Cor) images acquired on 2014 October 29 (Fig. 1, right panel)  outlines the existence of a broad coronal structure in the southeast region of the Sun that may be at the origin of the \tii\ emission.
In order to investigate the above scenario, we  assume for simplicity that the
CME-driven shock surface at any given time has the same spherical shape as the observed CME bubble at the same time, apart from a factor defining the standoff distance between CME and shock surface. 
In general, for a fast-moving CME, the overlying shock surface should be oblate, that is, nonspherical.
For this particular event however the overall dynamics of a fast-expanding sphere with a small translating component away from the Sun implies that the shock surface can be aptly approximated with a spherical shock surface enveloping the CME leading edge: when detected by AIA/SDO, shocks in their initial stage often do appear as projections of spherical surfaces on the plane of the sky (e.g., \citealt{Ma2011,Gopalswamy2012}).
The radius of curvature $r_{sh}$ of the shock surface was determined by assuming that $r_{sh} = a_0 + a_1r_c$, where $r_c$ is the radius of the EUV bubble and $a_0$, $a_1$ are free parameters. 
The angle of propagation of the CME bubble was also left as a free parameter.
We further supposed the streamers to be straight and radial in the $y-x$ plane (i.e., as seen from the poles of the Sun), yielding two additional unknown angular positions.
Finally, with an iterative fitting procedure, we determined the five parameters that minimized the least-square difference between calculated and observed locations of the \tii\ radio sources in the $x-z$ plane.
In this way, we were able to infer both the time-evolving 3D geometry of the CME/shock and the spatial location of the axes of the two streamers (depicted as red and blue lines in Fig.~\ref{Fig4}).
The best match for the angle of propagation of the CME bubble was found at 15.3\dg\ towards the observer with respect to the plane of the sky (see Fig.~\ref{Fig4}), very near to the heliolongitude of the associated flare.
Figure~\ref{Fig5}a shows how the variation in the standoff distance between shock surface and CME leading edge depends on time (and thus distance): the standoff increases with time because the rate of expansion of the shock surface is higher than that of the underlying CME.
The expansion of the CME acts as a piston close to the Sun thus increasing the standoff distance and the shock speed since plasma material cannot flow behind the driver.
Once the piston loses energy and decelerates in its outward propagation, the shock is expected to detach and continue its propagation, albeit with no additional energy supply by the piston.

\section{Coronal magnetic field}

The upstream \alfvenic\ Mach number $M_{\rm A}$ and the magnetic field strength $B$ can be derived from the standoff distance $\Delta r \equiv r_{\rm sh} - r_{\rm cme}$ between the shock front $r_{\rm sh}$ and the CME leading edge $r_{\rm cme}$ by employing a model adapted by \citet{Gopalswamy2011} but originally developed for the bow shock of the Earth.
Physically, the shock Mach number $M$ and the radius of curvature $r_{\rm cme}$ of the nose of the driver control the standoff distance.
The ratio $\delta$ between $\Delta r$ and $r_{\rm cme}$ can be expressed as $\delta \equiv \Delta r/r_{\rm cme} = 0.81 [(\gamma - 1)M^2+2]/[(\gamma - 1)(M^2 - 1)],$ (e.g., \citealt{Farris1994}) where $\gamma$ is the adiabatic index (we assume $\gamma=4/3$ and $\gamma=5/3$ as two possible values).
This expression allows the applicability of the model to the low-$M$ regime provided that $M$ is substituted for the fast magnetosonic Mach number (\citealt{Fairfield2001}).
In the inner corona, $v_{\rm A}$ substantially exceeds the sound speed meaning that $M$ can be confidently replaced by $M_{\rm A}$.
It then follows that $M_{\rm A}^2=1+[\delta/0.81-(\gamma-1)/(\gamma+1)]^{-1},$
implying that $M_{\rm A}$ (and thus the coronal magnetic field strength $B$, given the upstream electron density profile $n_{\rm e}$) can be obtained at any given time by measuring $\delta$.
We point out that the above method is based on a model by \citet{Farris1994}, further extended by \citet{Russell2002}, which was specifically designed for the bow shock of the terrestrial magnetosphere. 
As such, the evaluation of the standoff distance in that model is intended to be strictly valid for the nose of a CME. 
It is then important to discuss how appropriate such a model is for a coronal CME-driven shock, for which, moreover, as in our case, the standoff distance is calculated not at the nose but at the flanks of the shock.
In order to check the validity of the model in the coronal environment, Schmidt et al. (2016) used a 3D magnetohydrodynamic code to test the accuracy of the \citet{Gopalswamy2011} method by applying it to a simulation of a real CME event, where the standoff distances between the driven shock and the CME could be determined within the simulation box. 
As a result, the predicted magnetic field strength profile agreed very well (within 30\% between 1.8 and 10 \rsun) with the magnetic field strength profile found in the simulation domain, thus providing strong support for the \citet{Gopalswamy2011} method in a coronal environment. 
As for the second issue, that is, the reliability of the model when the standoff distance is evaluated at the flanks of the shock surface, it is obvious that in the general case of a fast moving CME the shock surface would be better represented by a 3D oblate surface. 
In that case, our model would yield unreliable results away from the CME nose. 
In our specific case however, since the translational motion of the CME away from the Sun is negligible with respect to its rapidly expanding motion in all directions, we posit the standoff distance between the leading edge of the CME  and the piston-driven shock surface to be uniform at all angles, at least in the inner corona where our measurements were taken. 
We thus claim, with no further theoretical justification, that the above model, which would be strictly valid only at the nose of the shock, is also applicable to its flanks.

By applying the above theoretical framework, we estimated the pre-shock $v_{\rm A}$ and $M_{\rm A}$ (Fig.~\ref{Fig5}b) along the two streamers (the colors of the curves correspond to the colors of the streamers in Fig.~\ref{Fig4}).
Before declining away from the Sun, $v_{\rm A}$ reached a maximum of $800-900$ \kms\ at about $1.2-1.3$ \rsun\ along the two streamers.
Indeed, the \tii\ ignites at the height, around 13:08:12 UT, when $v_{\rm sh}$ (and thus $M_{\rm A}$) was high enough (about 1.16).
As seen in the graph, variations due to different values for $\gamma$ are not important in the following discussion.
Since $v_A = v_{sh}/M_A$ (the solar wind speed at low coronal heights is negligible when compared to $v_{\rm sh}$), we are finally able to derive the magnetic field strength as $B = 4.92\cdot10^{-7}  {v_{\rm A}} \sqrt{n_{\rm e}}  ~~ [{\rm G}],$ where $n_{\rm e}$ is the upstream electron density in [cm$^{-3}$].
In this expression, all parameters are known as a function of heliocentric distance $r$ (the radial dependence of $n_{\rm e}$ is obtained from the emission frequency of the \tii\ radio burst and its location as inferred from the images of NRH by fitting a \cite{Newkirk1961}-like function of the type $n_{\rm e}(r)=\alpha\cdot10^{\beta/r}$; see Fig.~\ref{Fig5}c).
We thus obtain two estimates of the magnetic field strength profile between 1.2 and 1.4 \rsun\ according to the plasma properties overlying the appropriate streamers interaction with the expanding shock surface.
Figure~\ref{Fig5}d compares the magnetic field radial profile as derived from this work with the profile obtained in the inner heliosphere by \cite{Mancuso2013b} together with a number of near-Sun estimates of $B$ based on different methods.
A simple radial profile represented by a power law of the form $B(r) = \alpha r^\beta$ [G] with $\alpha = 12.6 \pm 2.5$ and $\beta = -4$ nicely fits previous estimates between 1.1 and 2.0 \rsun\ from the center of the Sun.

\section{Discussion and conclusions}

The origin of band-splitting in \tii\ radio bursts is often interpreted as being due to simultaneous emission from plasma upstream and downstream of a propagating shock front (e.g., \citealt{Chrysaphi2018,Zucca2018,Frassati2019}). 
Under this hypothesis, the upstream magnetic field strength $B$ can be promptly estimated by inferring the density jump at the shock front from the instantaneous band split.
Band-splitting, however, may also originate from the intersection of the expanding shock surface with adjacent low-\alfven-speed coronal streamer structures.
The key result of this Letter is that, under this latter assumption, it is still possible to reliably infer the magnetic field strength $B$ and its profile in the inner corona when favorable conditions are met, even without the use of stereoscopic observations.

The band-splitted \tii\ radio burst analyzed in this work was associated with a CME-driven shock that occurred on 2014 October 30 over the southeast limb of the Sun.
The CME dynamics was inferred from EUV observations obtained with AIA/SDO in the 211 \AA\ passband.
The CME-driven shock was modeled as a spherical expanding/translating surface, whose radius of curvature was determined through a least-square fitting procedure to identify the best match between the 3D model and the observational constraints given by the spatiotemporal evolution of the \tii\ radio sources (assumed to be emitted at the intersection of the shock surface with two adjacent low-\alfven\ speed coronal streamer structures), whose positions were imaged in a wide frequency range covered by the NRH.
From the resulting derived profile of the standoff distance between shock and  leading edge of the CME, we were finally able to infer, by employing a method adapted by \citet{Gopalswamy2011}, the pre-shock magnetic field strength in the inner corona between 1.2 and 1.4 \rsun\ from the center of the Sun.
When combined together with previous estimates, the radial profile of the magnetic field strength can be represented by a power law of the form $B(r) = (12.6 \pm 2.5) r^{-4}$ [G] in the heliocentric distance range $1.1-2.0$ \rsun.

We finally point out that evidence for a shock was also found in AIA data near the angular position where the \tii\ emission was excited.
Differential emission measure analysis of EUV data will allow us to infer both the compression ratio and the temperature jump at the shock front (Frassati et al. in preparation).

\begin{acknowledgements} 
The authors would like to thank the anonymous referee for the valuable comments which improved the manuscript. 
We thank the teams of e-Callisto, NRH, and SDO/AIA for their open-data use policy. 
K-Cor data are courtesy of MLSO, operated by HAO, as part of NCAR, supported by the NSF.
F. F. acknowledges support from INAF Ph.D. Grant.
\end{acknowledgements}

\bibliographystyle{aa}
\bibliography{biblio}

\begin{thebibliography}{28}
\expandafter\ifx\csname natexlab\endcsname\relax\def\natexlab#1{#1}\fi

\bibitem[{{Aurass} {et~al.}(1987){Aurass}, {Kurths}, {Mann}, {Chernov}, \&
  {Karlicky}}]{Aurass1987}
{Aurass}, H., {Kurths}, J., {Mann}, G., {Chernov}, G.~P., \& {Karlicky}, M.
  1987, \solphys, 108, 131

\bibitem[{{Bemporad} {et~al.}(2016){Bemporad}, {Susino}, {Frassati}, \&
  {Fineschi}}]{Bemporad2016}
{Bemporad}, A., {Susino}, R., {Frassati}, F., \& {Fineschi}, S. 2016, Frontiers
  in Astronomy and Space Sciences, 3, 17

\bibitem[{{Benz} {et~al.}(2005){Benz}, {Monstein}, \& {Meyer}}]{Benz2005}
{Benz}, A.~O., {Monstein}, C., \& {Meyer}, H. 2005, \solphys, 226, 143

\bibitem[{{Chrysaphi} {et~al.}(2018){Chrysaphi}, {Kontar}, {Holman}, \&
  {Temmer}}]{Chrysaphi2018}
{Chrysaphi}, N., {Kontar}, E.~P., {Holman}, G.~D., \& {Temmer}, M. 2018, \apj,
  868, 79

\bibitem[{{Fairfield} {et~al.}(2001){Fairfield}, {Cairns}, {Desch}, {Szabo},
  {Lazarus}, \& {Aellig}}]{Fairfield2001}
{Fairfield}, D.~H., {Cairns}, I.~H., {Desch}, M.~D., {et~al.} 2001, \jgr, 106,
  25361

\bibitem[{{Farris} \& {Russell}(1994)}]{Farris1994}
{Farris}, M.~H. \& {Russell}, C.~T. 1994, \jgr, 99, 17

\bibitem[{{Frassati} {et~al.}(2019){Frassati}, {Susino}, {Mancuso}, \&
  {Bemporad}}]{Frassati2019}
{Frassati}, F., {Susino}, R., {Mancuso}, S., \& {Bemporad}, A. 2019, \apj, 871,
  212

\bibitem[{{Gopalswamy} {et~al.}(2012){Gopalswamy}, {Nitta}, {Akiyama},
  {M{\"a}kel{\"a}}, \& {Yashiro}}]{Gopalswamy2012}
{Gopalswamy}, N., {Nitta}, N., {Akiyama}, S., {M{\"a}kel{\"a}}, P., \&
  {Yashiro}, S. 2012, \apj, 744, 72

\bibitem[{{Gopalswamy} \& {Yashiro}(2011)}]{Gopalswamy2011}
{Gopalswamy}, N. \& {Yashiro}, S. 2011, \apjl, 736, L17

\bibitem[{{Kerdraon} \& {Delouis}(1997)}]{Kerdraon1997}
{Kerdraon}, A. \& {Delouis}, J.-M. 1997, in Lecture Notes in Physics, Berlin
  Springer Verlag, Vol. 483, Coronal Physics from Radio and Space Observations,
  ed. G.~{Trottet}, 192

\bibitem[{{Lemen} {et~al.}(2012){Lemen}, {Title}, {Akin}, {Boerner}, {Chou},
  {Drake}, {Duncan}, {Edwards}, {Friedlaender}, {Heyman}, {Hurlburt}, {Katz},
  {Kushner}, {Levay}, {Lindgren}, {Mathur}, {McFeaters}, {Mitchell}, {Rehse},
  {Schrijver}, {Springer}, {Stern}, {Tarbell}, {Wuelser}, {Wolfson}, {Yanari},
  {Bookbinder}, {Cheimets}, {Caldwell}, {Deluca}, {Gates}, {Golub}, {Park},
  {Podgorski}, {Bush}, {Scherrer}, {Gummin}, {Smith}, {Auker}, {Jerram},
  {Pool}, {Soufli}, {Windt}, {Beardsley}, {Clapp}, {Lang}, \&
  {Waltham}}]{Lemen2012}
{Lemen}, J.~R., {Title}, A.~M., {Akin}, D.~J., {et~al.} 2012, \solphys, 275, 17

\bibitem[{{Lin} {et~al.}(2000){Lin}, {Penn}, \& {Tomczyk}}]{Lin2000}
{Lin}, H., {Penn}, M.~J., \& {Tomczyk}, S. 2000, \apjl, 541, L83

\bibitem[{{Ma} {et~al.}(2011){Ma}, {Raymond}, {Golub}, {Lin}, {Chen}, {Grigis},
  {Testa}, \& {Long}}]{Ma2011}
{Ma}, S., {Raymond}, J.~C., {Golub}, L., {et~al.} 2011, \apj, 738, 160

\bibitem[{{Mancuso} \& {Garzelli}(2013{\natexlab{a}})}]{Mancuso2013a}
{Mancuso}, S. \& {Garzelli}, M.~V. 2013{\natexlab{a}}, \aap, 560, L1

\bibitem[{{Mancuso} \& {Garzelli}(2013{\natexlab{b}})}]{Mancuso2013b}
{Mancuso}, S. \& {Garzelli}, M.~V. 2013{\natexlab{b}}, \aap, 553, A100

\bibitem[{{Mancuso} \& {Raymond}(2004)}]{Mancuso2004}
{Mancuso}, S. \& {Raymond}, J.~C. 2004, \aap, 413, 363

\bibitem[{{Mancuso} \& {Spangler}(2000)}]{Mancuso2000}
{Mancuso}, S. \& {Spangler}, S.~R. 2000, \apj, 539, 480

\bibitem[{{Nakariakov} \& {Ofman}(2001)}]{Nakariakov2001}
{Nakariakov}, V.~M. \& {Ofman}, L. 2001, \aap, 372, L53

\bibitem[{{Newkirk}(1961)}]{Newkirk1961}
{Newkirk}, Jr., G. 1961, \apj, 133, 983

\bibitem[{{Ofman} \& {Wang}(2008)}]{Ofman2008}
{Ofman}, L. \& {Wang}, T.~J. 2008, \aap, 482, L9

\bibitem[{{Patsourakos} {et~al.}(2010){Patsourakos}, {Vourlidas}, \&
  {Kliem}}]{Patsourakos2010}
{Patsourakos}, S., {Vourlidas}, A., \& {Kliem}, B. 2010, \aap, 522, A100

\bibitem[{{Riddle}(1974)}]{Riddle1974}
{Riddle}, A.~C. 1974, \solphys, 35, 153

\bibitem[{{Russell} \& {Mulligan}(2002)}]{Russell2002}
{Russell}, C.~T. \& {Mulligan}, T. 2002, \planss, 50, 527

\bibitem[{{Smerd} {et~al.}(1975){Smerd}, {Sheridan}, \& {Stewart}}]{Smerd1975}
{Smerd}, S.~F., {Sheridan}, K.~V., \& {Stewart}, R.~T. 1975, \aplett, 16, 23

\bibitem[{{Spangler}(2005)}]{Spangler2005}
{Spangler}, S.~R. 2005, \ssr, 121, 189

\bibitem[{{van Haarlem} {et~al.}(2013){van Haarlem}, {Wise}, {Gunst}, {Heald},
  {McKean}, {Hessels}, {de Bruyn}, {Nijboer}, {Swinbank}, {Fallows},
  {Brentjens}, {Nelles}, {Beck}, {Falcke}, {Fender}, {H{\"o}randel},
  {Koopmans}, {Mann}, {Miley}, {R{\"o}ttgering}, {Stappers}, {Wijers},
  {Zaroubi}, {van den Akker}, {Alexov}, {Anderson}, {Anderson}, {van Ardenne},
  {Arts}, {Asgekar}, {Avruch}, {Batejat}, {B{\"a}hren}, {Bell}, {Bell}, {van
  Bemmel}, {Bennema}, {Bentum}, {Bernardi}, {Best}, {B{\^i}rzan}, {Bonafede},
  {Boonstra}, {Braun}, {Bregman}, {Breitling}, {van de Brink}, {Broderick},
  {Broekema}, {Brouw}, {Br{\"u}ggen}, {Butcher}, {van Cappellen}, {Ciardi},
  {Coenen}, {Conway}, {Coolen}, {Corstanje}, {Damstra}, {Davies}, {Deller},
  {Dettmar}, {van Diepen}, {Dijkstra}, {Donker}, {Doorduin}, {Dromer}, {Drost},
  {van Duin}, {Eisl{\"o}ffel}, {van Enst}, {Ferrari}, {Frieswijk}, {Gankema},
  {Garrett}, {de Gasperin}, {Gerbers}, {de Geus}, {Grie{\ss}meier}, {Grit},
  {Gruppen}, {Hamaker}, {Hassall}, {Hoeft}, {Holties}, {Horneffer}, {van der
  Horst}, {van Houwelingen}, {Huijgen}, {Iacobelli}, {Intema}, {Jackson},
  {Jelic}, {de Jong}, {Juette}, {Kant}, {Karastergiou}, {Koers}, {Kollen},
  {Kondratiev}, {Kooistra}, {Koopman}, {Koster}, {Kuniyoshi}, {Kramer},
  {Kuper}, {Lambropoulos}, {Law}, {van Leeuwen}, {Lemaitre}, {Loose}, {Maat},
  {Macario}, {Markoff}, {Masters}, {McFadden}, {McKay-Bukowski}, {Meijering},
  {Meulman}, {Mevius}, {Middelberg}, {Millenaar}, {Miller-Jones}, {Mohan},
  {Mol}, {Morawietz}, {Morganti}, {Mulcahy}, {Mulder}, {Munk}, {Nieuwenhuis},
  {van Nieuwpoort}, {Noordam}, {Norden}, {Noutsos}, {Offringa}, {Olofsson},
  {Omar}, {Orr{\'u}}, {Overeem}, {Paas}, {Pandey-Pommier}, {Pandey}, {Pizzo},
  {Polatidis}, {Rafferty}, {Rawlings}, {Reich}, {de Reijer}, {Reitsma},
  {Renting}, {Riemers}, {Rol}, {Romein}, {Roosjen}, {Ruiter}, {Scaife}, {van
  der Schaaf}, {Scheers}, {Schellart}, {Schoenmakers}, {Schoonderbeek},
  {Serylak}, {Shulevski}, {Sluman}, {Smirnov}, {Sobey}, {Spreeuw}, {Steinmetz},
  {Sterks}, {Stiepel}, {Stuurwold}, {Tagger}, {Tang}, {Tasse}, {Thomas},
  {Thoudam}, {Toribio}, {van der Tol}, {Usov}, {van Veelen}, {van der Veen},
  {ter Veen}, {Verbiest}, {Vermeulen}, {Vermaas}, {Vocks}, {Vogt}, {de Vos},
  {van der Wal}, {van Weeren}, {Weggemans}, {Weltevrede}, {White}, {Wijnholds},
  {Wilhelmsson}, {Wucknitz}, {Yatawatta}, {Zarka}, {Zensus}, \& {van
  Zwieten}}]{vanHarlem2013}
{van Haarlem}, M.~P., {Wise}, M.~W., {Gunst}, A.~W., {et~al.} 2013, \aap, 556,
  A2

\bibitem[{{Verwichte} {et~al.}(2009){Verwichte}, {Aschwanden}, {Van
  Doorsselaere}, {Foullon}, \& {Nakariakov}}]{Verwichte2009}
{Verwichte}, E., {Aschwanden}, M.~J., {Van Doorsselaere}, T., {Foullon}, C., \&
  {Nakariakov}, V.~M. 2009, \apj, 698, 397

\bibitem[{{Zucca} {et~al.}(2018){Zucca}, {Morosan}, {Rouillard}, {Fallows},
  {Gallagher}, {Magdalenic}, {Klein}, {Mann}, {Vocks}, {Carley}, {Bisi},
  {Kontar}, {Rothkaehl}, {Dabrowski}, {Krankowski}, {Anderson}, {Asgekar},
  {Bell}, {Bentum}, {Best}, {Blaauw}, {Breitling}, {Broderick}, {Brouw},
  {Br{\"u}ggen}, {Butcher}, {Ciardi}, {de Geus}, {Deller}, {Duscha},
  {Eisl{\"o}ffel}, {Garrett}, {Grie{\ss}meier}, {Gunst}, {Heald}, {Hoeft},
  {H{\"o}randel}, {Iacobelli}, {Juette}, {Karastergiou}, {van Leeuwen},
  {McKay-Bukowski}, {Mulder}, {Munk}, {Nelles}, {Orru}, {Paas}, {Pandey},
  {Pekal}, {Pizzo}, {Polatidis}, {Reich}, {Rowlinson}, {Schwarz}, {Shulevski},
  {Sluman}, {Smirnov}, {Sobey}, {Soida}, {Thoudam}, {Toribio}, {Vermeulen},
  {van Weeren}, {Wucknitz}, \& {Zarka}}]{Zucca2018}
{Zucca}, P., {Morosan}, D.~E., {Rouillard}, A.~P., {et~al.} 2018, \aap, 615,
  A89

\end{thebibliography}

\end{document}